\documentclass[graybox]{svmult}

\usepackage{mathptmx}       
\usepackage{helvet}         
\usepackage{courier}        
\usepackage{type1cm}        
\usepackage{wrapfig}                            
\usepackage{makeidx}         
\usepackage{graphicx}        
\usepackage{amsmath}                             
\usepackage{multicol}        
\usepackage[bottom]{footmisc}

\makeindex             

\begin{document}

\title*{Analysis of Ultra High Energy Muons at the INO-ICAL Using Pair-Meter Technique}
\author{Jaydip Singh, Srishti Nagu and Jyotsna Singh}
\institute{Jaydip Singh, Srishti Nagu and Dr. Jyotsna Singh \at Lucknow University, Department of Physics,Lucknow-226007,India \email{jaydip.singh@gmail.com, srishtinagu19@gmail.com
and jyo2210@yahoo.com}}
%
%
\maketitle
\vspace{-25mm}
\abstract{The proposed ICAL detector at INO is a large sized 
underground magnetized iron detector. ICAL is designed to reconstruct muon momentum using 
magnetic spectrometers. Energy measurement using magnets fail for muons in TeV range, since the angular deflection of the muon 
in the magnetic field is
negligible and the muon tracks become nearly straight. A new technique for measuring the energy of muons in the TeV range is used by the CCFR\cite{6} 
neutrino detector, known as the
Pair-Meter technique.This technique estimates muon energy from measurements of the energy deposited by the muon in many layers of an 
iron-calorimeter through 
e$^+$ and e$^-$ pair production. In this work we have performed Geant4 based preliminary analysis for iron plates and have demonstrated the 
observational feasibility of very 
high energy muons (1TeV-1000TeV) in a large mass underground detector operating as a pair-meter. This wide range of energy spectrum 
will be helpful for 
studying the cosmic rays in the Knee region and an understanding of the atmospheric neutrino flux for the present and future ultra high-energy 
atmospheric neutrino experiments.}

\section{Introduction}
 ICAL at India-based Neutrino Observatory(INO) is a 52 ktons Iron CALorimeter(ICAL) detector\cite{5} proposed to be built at Theni district 
 of Tamilnadu in Southern India. It is designed primarily to study the atmospheric neutrino flavor oscillations through interaction of atmospheric neutrinos.
 The main goal of the ICAL detector is to make precise measurements of neutrino oscillation parameters and determine the neutrino mass hierarchy\cite{5}.
 At a depth of around 
 1.2km underground, the INO-ICAL detector will be the biggest magnetized detector to measure cosmic ray muon flux with the capability to distinguish $\mu$+ from 
 $\mu$-. The existing direct and indirect method of muon spectrometry in experiments at accelerators and in cosmic rays(magnetic spectrometers, transition 
 radiation detectors) involve grave technical problems and fundamental limitations in the energy region $\geq$ 10$^{13}$ eV. These disadvantages are absent in 
 the method for estimating the muon energy by energy of secondary cascade formed by muons in thick layers of matter mainly due to the process of direct production 
 of  e$^+$ and e$^-$ pairs. Using this technique we can observe very high energy muons (1 TeV-1000 TeV) in a large mass underground detector operating as a 
 pair-meter at INO-ICAL. This energy range corresponds to primary cosmic rays which are approximately in 50TeV-50PeV energy range.\cite{1}.
 
This work presents a simulation based on Geant4 for momentum reconstruction of muons in GeV energy range using INO-ICAL magnet and the latest version of 
Geant based INO-ICAL code, developed by the INO-collaboration. We have developed a separate Geant4 code for 
counting the muon bursts in iron plates for ultra high energy muon analysis using pair-meter techniques. Details of the  components and dimensions are 
discussed in ref \cite{5}.
 
 \begin{figure}[t]
\sidecaption[t]
\vspace{-3mm}
\includegraphics[scale=.75]{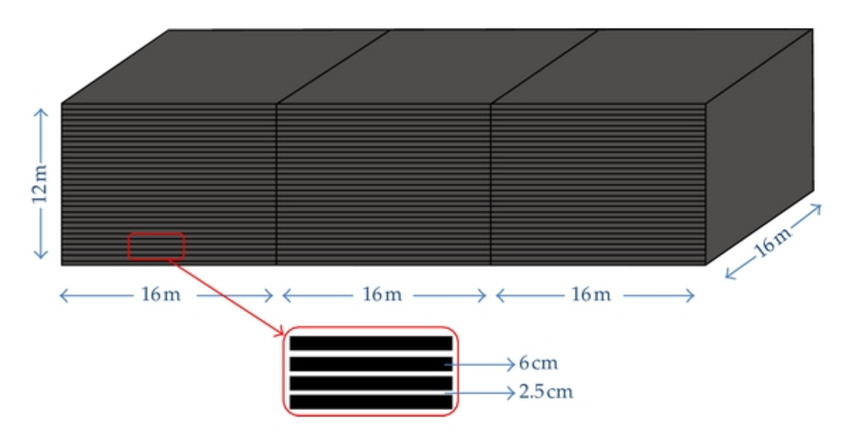}
%
%
\caption{Schematic view of three modules Iron Calorimeter detector proposed for INO. }
\label{fig:2}       
\end{figure}
\vspace{-3mm}
\section{Momentum Reconstruction analysis with ICAL Magnet:}
 In this section we have discussed the simulation for momentum reconstruction of muons using magnetic field. Details of the detector simulation for muons with energy of 
 few 10's of GeV 
 with older version of INO-ICAL code is already published in the ref\cite{4}. For simulating the response of high energy(100's of GeV) muons in the ICAL 
 detector, 10000 muons were propagated uniformly from a vertex randomly located inside 8m $\times$ 8m $\times$ 10m volume. This is the central region of the 
 central  module where the magnetic field is uniform of 1.5T. In our analysis we have considered only those events whose z coordinate of the input vertex lie 
 within z$_{in}$ $\leq$ 400cm which comprises the vertex to the central region. The input momentum and zenith angle are kept fixed in each case while the 
 azimuthal angle is uniformly averaged over the entire range -$\pi$ $\leq$ $\phi$ $\leq$ $\pi$. In each case, we study the number of reconstructed tracks, the 
 direction resolution, including up/down discrimination, and the zenith angle resolution. In this work we have followed the same approach as in ref \cite{4} for 
 muon response analysis up-to energies 500 GeV inside the detector. Momentum reconstruction efficiency in the energy range of 1-400 GeV is shown in Fig. 2 and 
 this energy range at the detector corresponds to the surface muon lying in the energy range 1600-2000 GeV from the top of the surface. Muon will lose 
 around 1600 GeV in the rock overburden to reach at the detector from the top surface. Energetic muons from other directions will also hit the detector since the 
 rock cover in other direction of detector is very huge, so we have not discussed it here. 
 
 \begin{figure}
\sidecaption
\includegraphics[scale=.55]{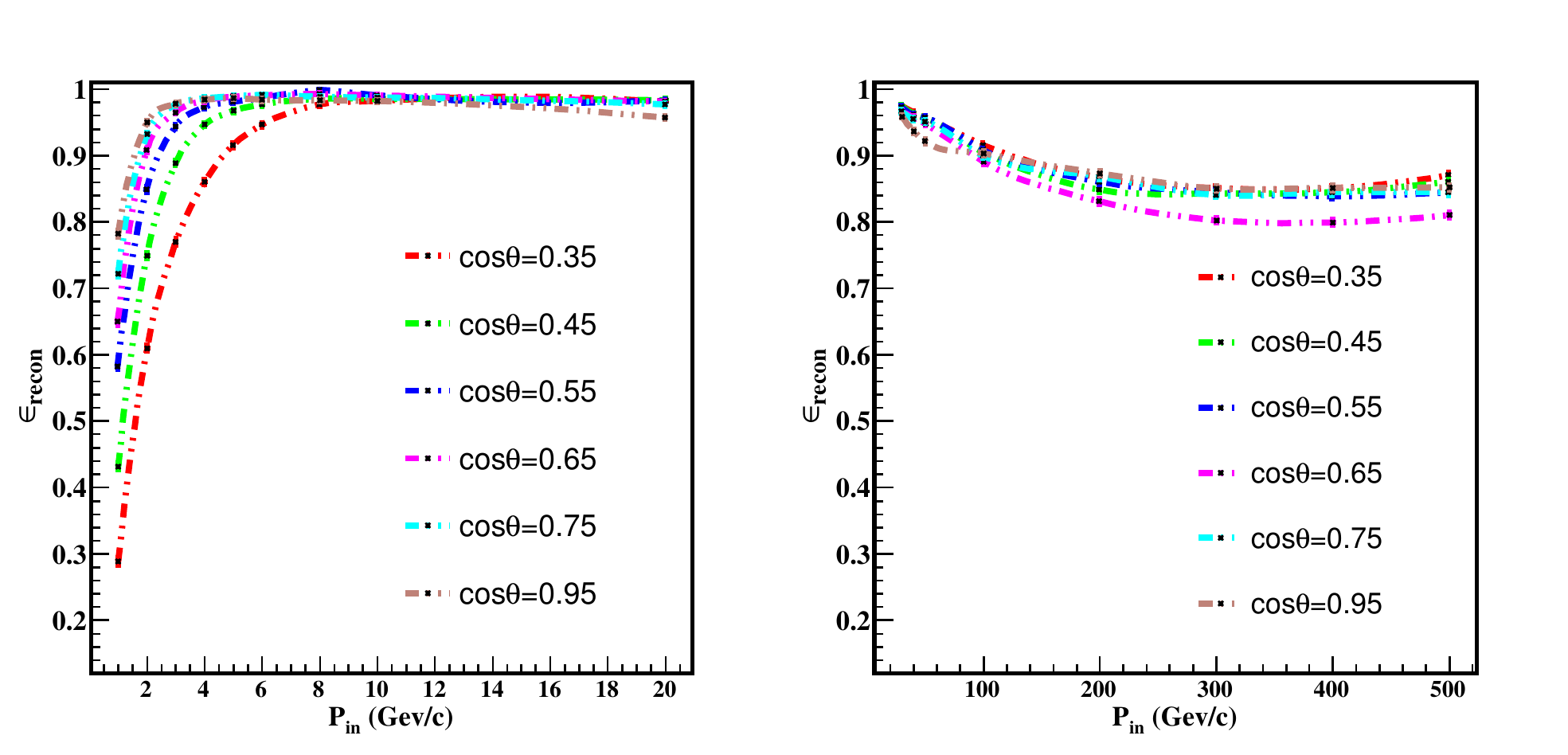}
%
%
\caption{The relative charge identification efficiency as a function of the input momentum for different cos$\theta$ values at low and high energy}
\label{fig:2}       
\end{figure}
  
  The momentum reconstruction efficiency($\epsilon_{rec}$) is defined as the ratio of the number of reconstructed events, n$_{rec}$, to the total number 
of generated events, N$_{total}$. We have \\

   \hspace{5cm}    $\epsilon_{rec}$ = $\boldmath\frac{{n_{rec}}}{\boldmath N_{total}}$ , \\
                  
   \hspace{3cm}  with error, $\delta \epsilon_{rec}$=$\boldmath\sqrt{(\epsilon_{rec}(1-\epsilon_{rec})/\boldmath N_{total})}$. \\
   
Figure 2 shows the muon momentum reconstruction efficiency as a function of input momentum for different cos$\theta$ bins in which left and right figures 
demonstrate detector response for low and high energy(in 100's of GeV) muon momentum respectively. One can see that the momentum reconstruction efficiency 
depends on the incident particle momentum, the angle of propagation and the strength of the magnetic field. As the input momentum increases, the reconstruction 
efficiency increases for all angles because with increase in energy the particle crosses more number of layers producing more hits in the detector. At higher 
energies the reconstruction efficiency starts decreasing, since the muons travel nearly straight without being deflected through the magnetic field of the detector. 
Track reconstruction is done using Kalman Filter techniques\cite{4}, tracks for few energies are plotted in the Fig.-3, which shows the deflected and undeflected
tracks depending on the energy of muons for fixed magnetic field(1.5T).

\begin{figure}
\sidecaption
\includegraphics[scale=.45]{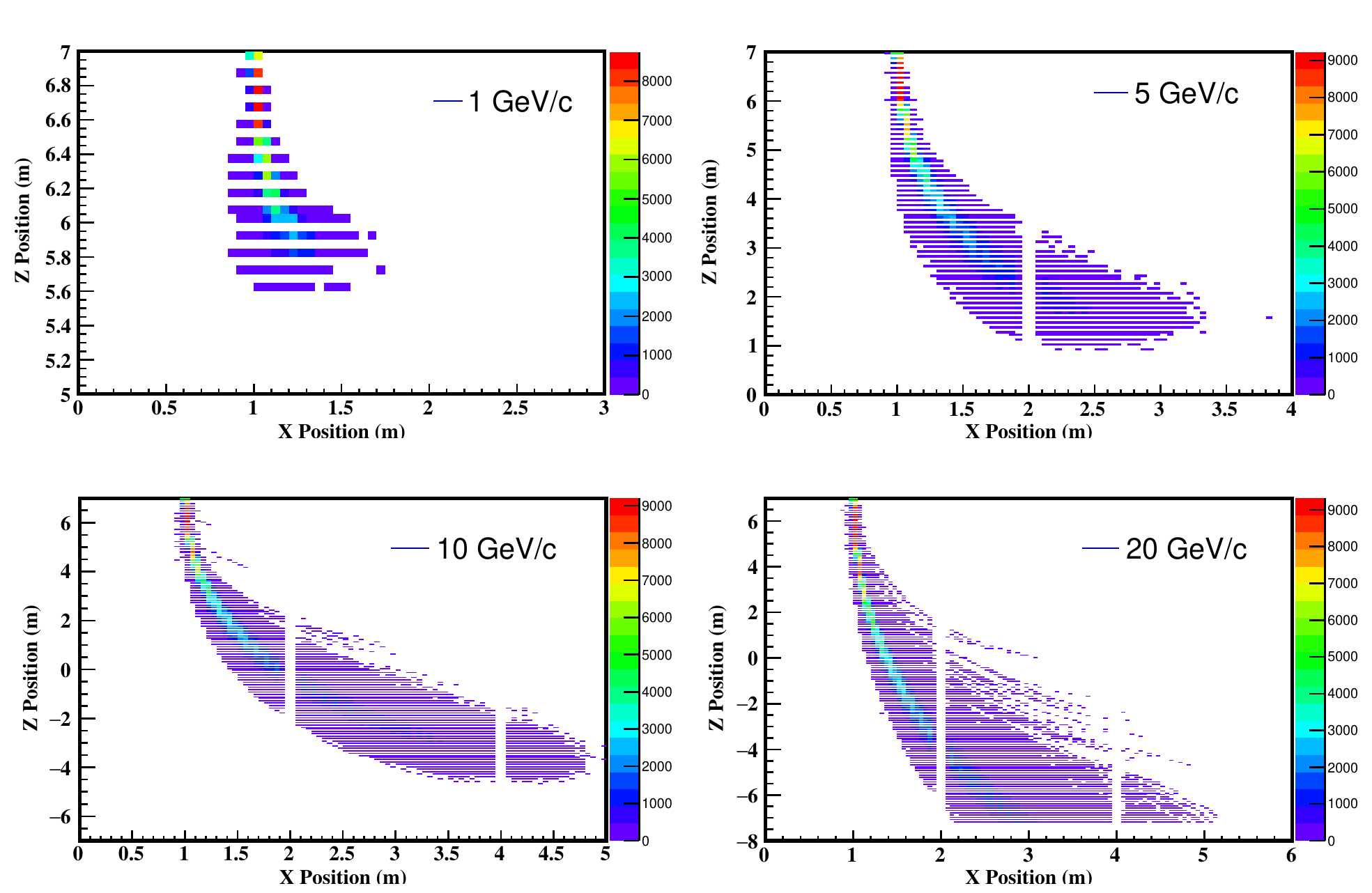}
\includegraphics[scale=.45]{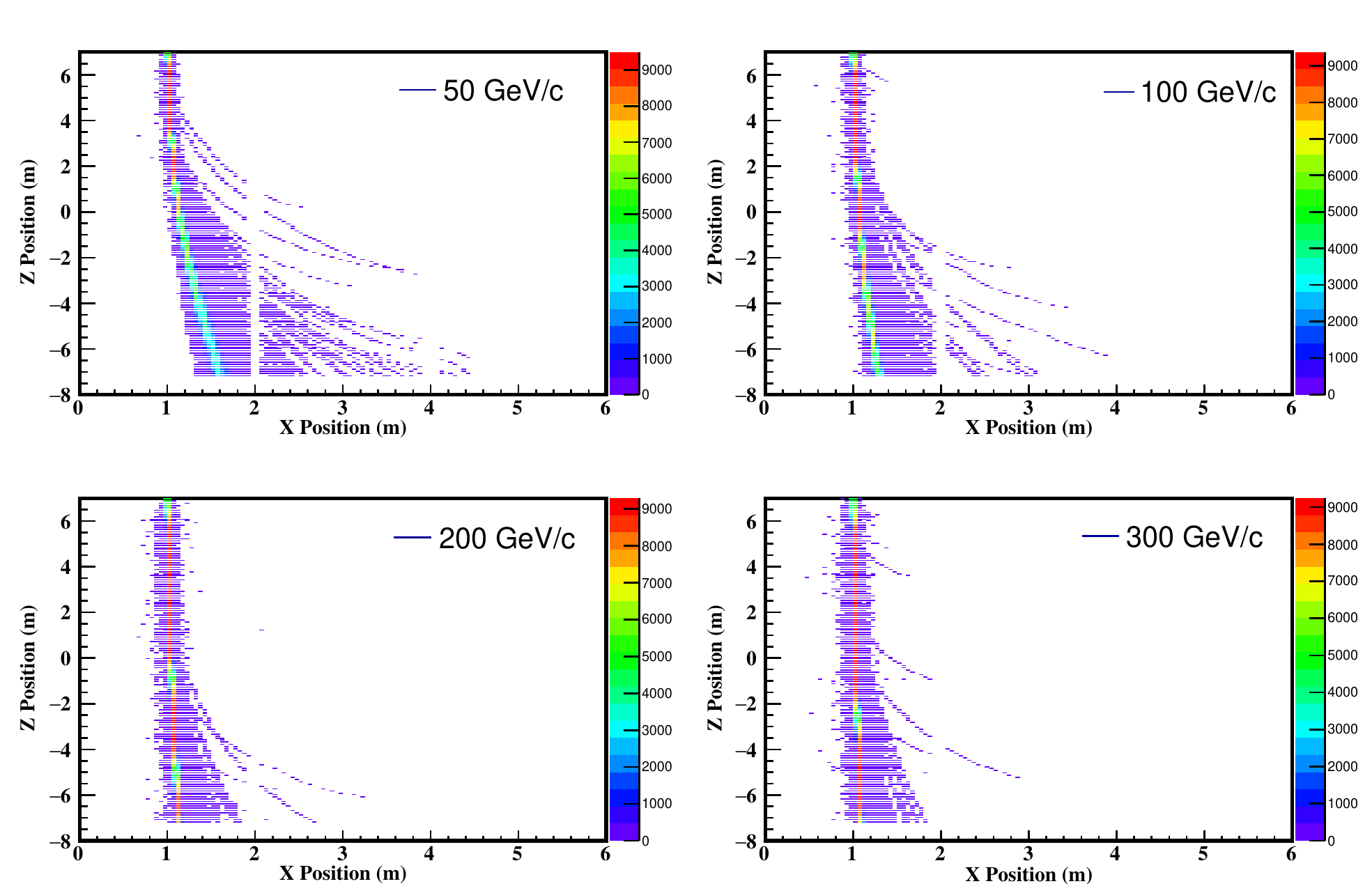}
%
%
\caption{Fixed energy muons track stored in X-Z plane of the detector going in the downward direction.}
\label{fig:1}       
\end{figure}

\paragraph{Pair Meter Techniques :} %
\begin{enumerate}
\item{High energy muons produce secondary cascades mainly due to electron pair production process.} \\
\item{It is one of the most important processes for muon interaction at TeV energies, pair creation cross section exceeds those of other muon interaction 
processes in a wide region of energy transfer:}\\
       \hspace{2cm}{100 MeV$\leq$ E$_{0}$ $\leq$ 0.1E$\mu$} , E$_{0}$ is Threshold energy.
\item Average energy loss for pair production increases linearly with muon energy, and in TeV region this process contributes over 50 percent of the total energy 
loss rate.
  
 \item Pair meter method for energy reconstruction of high energy muons has been used by the NuTeV/CCFR collaboration\cite{6}.  
\end{enumerate}

\subparagraph{Average No. of Burst Calculation\cite{1}:} The Cross section for e$^+$ and e$^-$ pair production by a muon with energy E$_{\mu}$ with energy 
transfer above a threshold E$_0$ grows as ln$^2$(2m$_e$ E$_{\mu}$/m$_{\mu}$E$_0$) where m$_{\mu}$ and m$_e$ are the muon and electron masses respectively. 
Defining v=E$_0$/E$_{\mu}$, above v$^{-1}$=10, this cross section dominates those for other muon energy loss process which generate observable cascades in its 
passage through dense matter.

\begin{itemize}
\item{The dependence of the pair meter production cross section on E$_{\mu}$ / E$_{0}$ then allows one to infer the muon energy by 
       counting the number of interaction cascade M in the detector with energies above a threshold E$_{0}$.
 \item Expression for differential pair production cross section is given by[1]:\\}
 
\begin{itemize}
\item{ \hspace{2cm} v$\dfrac{d\sigma}{dv}$ $\simeq$ $\dfrac{14\alpha}{9\pi t_{0}}$ln\bigg($\dfrac{km_{e}E_{\mu}}{E_{0}m_{\mu}}$\bigg) \hspace{3cm}   (1)}  \\
\item{where $\alpha$ = 1/137 ,  k $\simeq$1.8 and t$_{0}$ is the radiation length(r.l.) of the material , for 
     iron t$_{0}$ = 13.75 gm/cm$^{2}$ .}
\end{itemize}
\item{The Average number of interaction cascades M above a thereshold E$_{0}$ is given by :}

    \hspace{2cm}  M (E$_{0}$,E$_\mu$) = Tt$_{0}$ $\sigma$(E$_{0}$,E$_{\mu}$) \hspace{4cm}   (2) \\

      \hspace{2cm}  E$_\mu$ = ($E_{0}m_{\mu}/km_{e}$)exp($\sqrt{9\pi M / 7\alpha T - C}$)   \hspace{2cm}   (3) \ 

\begin{itemize}
 \item   where T is thickness of the target and $\sigma$(E$_{0}$,E$_{\mu}$) is the integrated cross section(in unit of 
   cm$^{2}$/gm)  and C $\simeq$ 1.4. ,\\
   
\hspace{2.5cm} $\sigma$(E$_{0}$,E$_{\mu}$) $\simeq$  $\dfrac{7\alpha}{9\pi t_{0}}$\Bigg(ln$^{2}$\bigg($\dfrac {km_{e} E_{\mu}}{E_{0}m_{\mu}}$\bigg)+C\Bigg) 
   \hspace{2cm}  (4)
\end{itemize}

\end{itemize}

%
\section{Counting the burst using Pair Meter:}
A muon traversing vertically from the top will cover 151$\times$5.6 $\simeq$ 845cm in iron plates,this corresponds to a path-length of $\simeq$  480 r.l.We can assume 
a average path-length for a muon of 450r.l.,therefore the number of cascades produced is given by equation (2) as shown in Fig. 4.
\label{sec:3}
\begin{figure}
\sidecaption
\includegraphics[scale=.35]{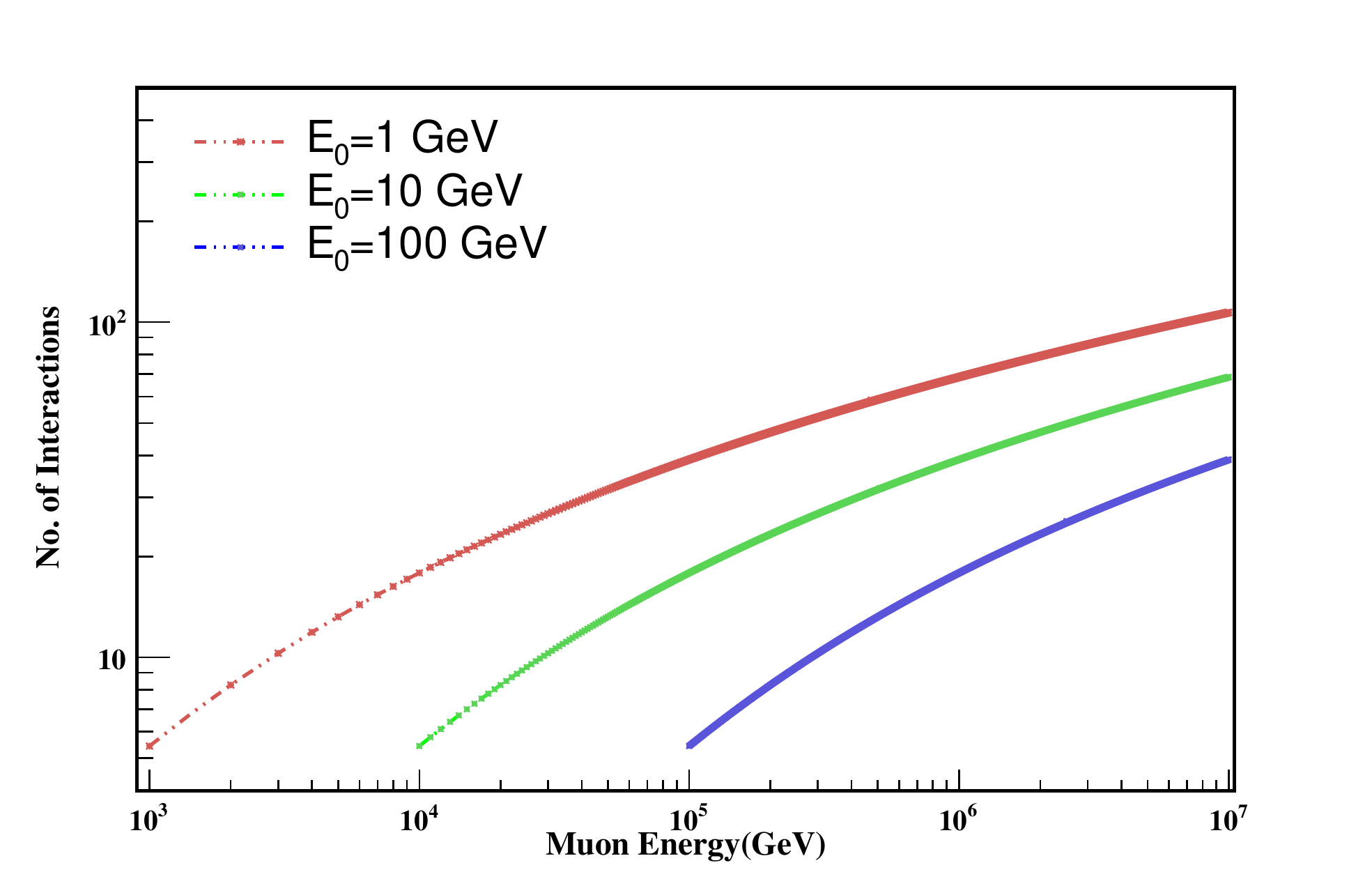}
%
%
\vspace{-5mm}
\caption{Average number of cascades above a threshold E$_0$ vs muon energy for E$_0$=1.0GeV, 10.0GeV and 100.00GeV, with T fixed to 450 r.l.}
\label{fig:2}       
\end{figure}
\subsection{Penetration Depth of Electron in the Iron Plates:} %
 Estimation of muons burst energy in iron plates will be evaluated by  e$^+$ and e$^-$ energy, for that electron produced in iron plates must come out of the iron plates 
 and hit the detector active element RPC (Resistive Plate Chamber). The energy loss of electron in iron is given by: E=E$_0$ e$^{-x/x_0}$, where x is distance traveled in the iron plate 
 and x$_0$ is the Radiation length. Electron of sufficient energy will come out of the iron plates and hit the RPC ,Fig.-5 shows electron energy and 
 their range. 

\begin{figure}
\sidecaption
\includegraphics[scale=.30]{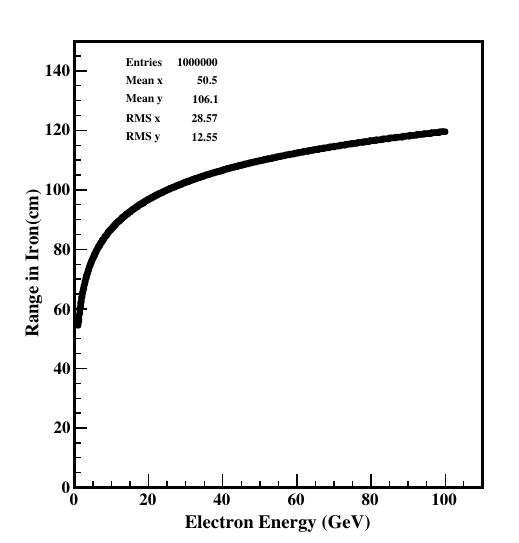}
%
%
\caption{ Energy of the electron and the corresponding penetrating range in the iron plates. }
\label{fig:2}       
\end{figure}

\subsubsection{10 TeV Muon Burst in the Iron Plates:}

A geant4 based code is developed for simulating the muons burst is iron plates, here horizontal axis is the z-axis of INO-ICAL detector, in which 152 layers 
of iron plates of width 5.6cm is placed vertically, interleaved with 4.0cm for placing the RPC.Muons are propagated using Geant4 particle generator class and 
generated bursts in the iron plates are counted, a 10TeV muon burst is shown in Fig.-6.
\begin{figure}
\sidecaption
\includegraphics[scale=.25]{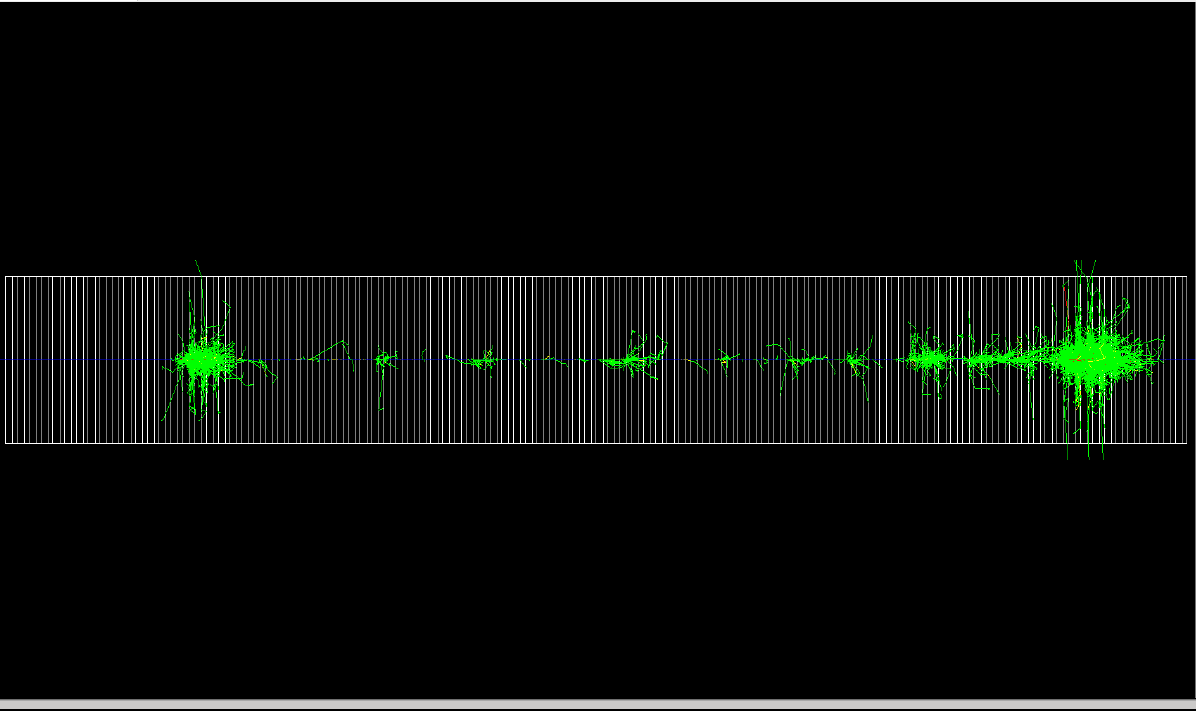}
%
%
\caption{Cascade Generation in the Iron Chamber , blue line(muon)represents z-axis of the detector and green line represents the electron-positron cascade in the x-y plane.}
\label{fig:1}       
\end{figure}

\subsection{Operating ICAL using Pair Meter Technique:} %
\vspace{-5mm}
\begin{figure}
\sidecaption
\includegraphics[scale=.40]{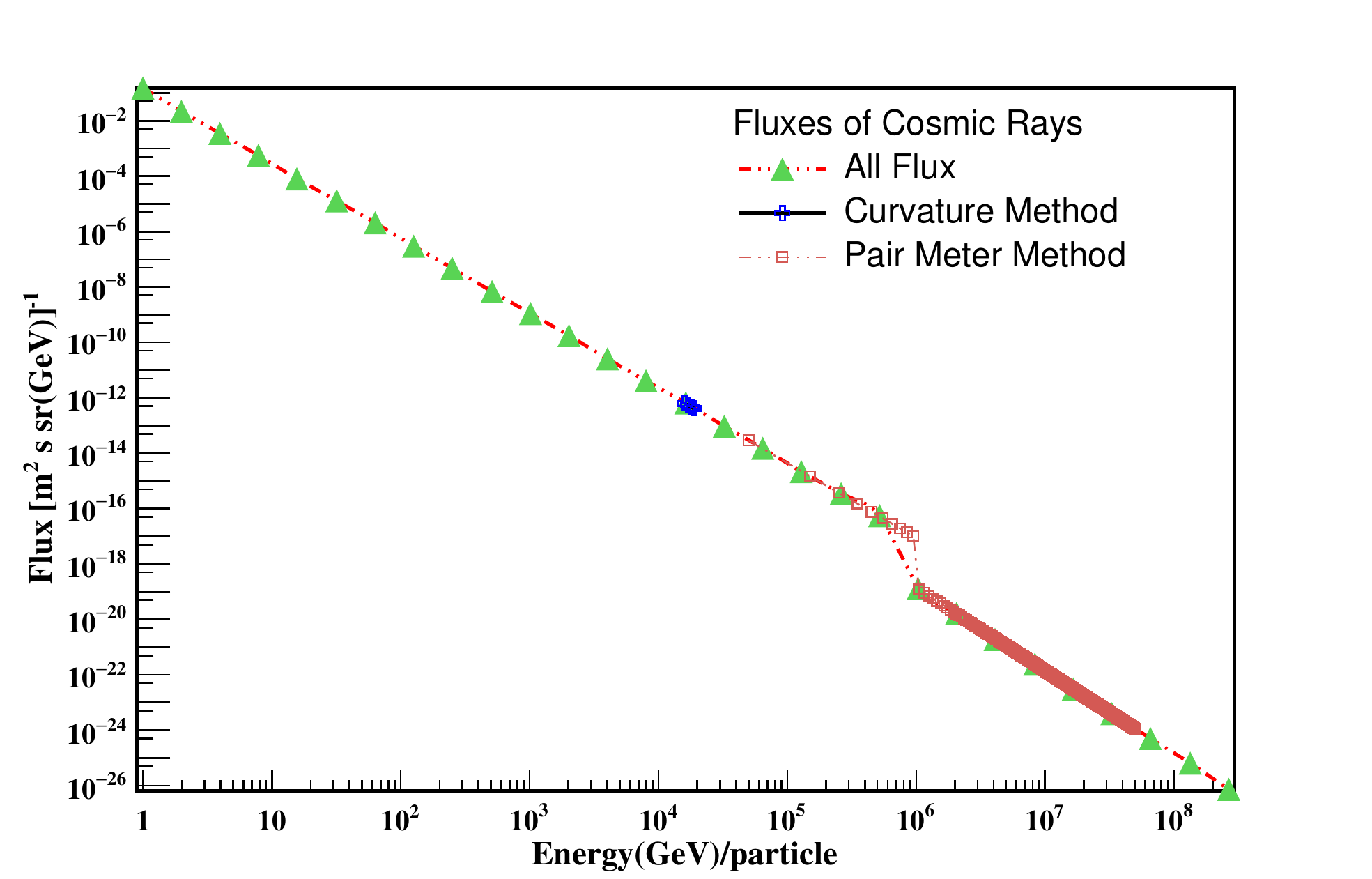}
%
%
\caption{Primary Cosmic Ray flux($\phi$ $\simeq$ KE$^{-\alpha}$, where $\alpha$ $\simeq$ 2.7 and for KNEE (3PeV)   $\alpha$ : 2.7 $\rightarrow$ 3) vs energy of 
primary particle, and limited range for ICAL to cover the spectrum using magnetic field and Pair Meter Technique.}
\label{fig:2}       
\end{figure}

 \section{Results and Discussion} 
  
      In section-2, we have discussed the limitation of magnetized ICAL detector to be used as a magnetic spectrometer, which limits the efficiency of the detector to
      discriminate between $\mu+$ and $\mu-$ at higher energies  and reconstruct momentum. Variation of momentum reconstruction efficiency as a function of 
      input momentum is shown in the Fig. 2, which shows clear fall in efficiency for energetic muons, we can also see the muon track in Fig. 3, which is
      undeflected for energetic muons. Finally we conclude that with ICAL detector we can do analysis for muons in the energy range of 1-400GeV. This corresponds to 
      surface muons in the energy range of 1600-2000GeV, because muons lose around 1600 Gev energy in the rock overburden to reach at the detector from the top of 
      the INO-ICAL surface. ICAL can not be used as a magnetic spectrometer for highly energetic cosmic ray muons. For energetic(TeV) muons, pair-meter 
      technique\cite{2} can be used for momentum reconstruction as discussed in section-3\cite{1}. This technique is tested by a few detectors,since INO-ICAL will be 
      large in dimensions so it will be a perfect machine to test the capability of this technique. We have developed a separate Geant4 code for counting the
      bursts in the iron plates and also a technique to measure the energy of the bursts with the produced electrons pairs in to the iron plates. In fig-6 we 
      can see the burst of muons in iron plate , some of the bursts are bigger and some are smaller as discussed in section -3.The variation of these 
      burst number is shown in fig. -4 , as a function of muon energy. 

\subsection{Summary and Conclusions:}
\begin{itemize}
 \item Pair meter technique can be used to measure the underground muon energy for an energy range of 1-1000 TeV at INO-ICAL detector operating as pair-meter.
 \item One can probe very high energy muon fluxes and primary cosmic rays in the Knee region , this will be helpful for estimation of the accurate background muon
  and neutrino fluxes for ultra high energy neutrino detectors.
 \item Our Geant4 analyses for central module of INO-ICAL detector is successfully performed and variation in the cascade number of varying energy is also observed 
  in the iron plates. Highly energetic bursts will be bigger in size and lesser energetic bursts will be smaller in size, in the X-Y plane of the iron plate. 
  
\end{itemize}

\begin{acknowledgement}
          This work is partially supported by Department of Physics, Lucknow University, Department of Atomic Energy, Harish-Chandra Research Institute, 
         Allahabad and INO collaboration. Financially it is supported by government of India, DST project no-SR/MF/PS02/2013, Department of Physics,Lucknow 
         University. We thank Prof. Raj Gandhi for useful discussion and providing hospitality to work in HRI,  Dr. Jyotsna Singh for her support and guidance
         in completing this work from Lucknow University. 
\end{acknowledgement}\vspace{-10mm}
%



\begin{thebibliography}{99.}%
%
%
\bibitem{1} Raj Gandhi , Sukanta Panda Journal-ref: JCAP 0607 (2006) 011 
%
\bibitem{2} R.P.KOKOULIN and A.A.PETRUKHIN , Nucl. Instruments and Method in Physics Research A263(1988)468-479 .
%
\bibitem{3} D.E. GROOM, N.V. MOKHOV, and S.STRIGANOV , Muon Stopping Power and Range .
%
\bibitem {4}A. Chatterjee et al.(INO Collab) JINST 9(2014) PO7001, July 2014

\bibitem{5}Shakeel Ahamed , M. Sajjad Atahar et al., "Physics Potential of the ICAL detector at the Indian Based Neutrino Observatory",(INO Collab) 
             arXiv:1505.07380
             
\bibitem{6} A.P. Chikkatur, L.Bugel et al. "Test of a Calorimetric technique for measuring the energy of Cosmic ray muon in TeV energy range", 
            Z. Phys. C 74, 279–289 (1997)  
             
\bibitem{7} GEANT4 collaboration, S. Agostinelli et al., GEANT4: a simulation toolkit, Nucl. Instrum. Meth. A
            506 (2003) 250; http://geant4.cern.ch/. 

\end{thebibliography}
\end{document}